\documentclass[lettersize,journal]{IEEEtran}
\IEEEoverridecommandlockouts

\usepackage{cite}
\usepackage{amsmath,amssymb,amsfonts}
\usepackage{breqn}
\usepackage[boxruled]{algorithm2e}
\usepackage{placeins}
\usepackage{algpseudocode}
\usepackage{graphicx}
\usepackage{textcomp}
\usepackage{xcolor}
\usepackage{float}
\usepackage{fancyhdr}
\def\BibTeX{{\rm B\kern-.05em{\sc i\kern-.025em b}\kern-.08em
    T\kern-.1667em\lower.7ex\hbox{E}\kern-.125emX}}

\begin{document}

\title{{Model-Free Control for Multi-Time Scale Dynamics of Grid-Connected Power Converters}}

\author{Dewan Mahnaaz Mahmud,~\IEEEmembership{Student member,~IEEE,} {Vinu Thomas},~\IEEEmembership{Member,~IEEE,} {Bogdan Marinescu},~\IEEEmembership{Member,~IEEE}

\thanks{Dewan Mahnaaz Mahmud and Vinu Thomas are with Nantes Université, École Centrale Nantes, LS2N, UMR 6004, F-44000 Nantes, France 
(email: dewan-mahnaaz.mahmud@ec-nantes.fr; vinu.thomas@ec-nantes.fr).}%
\thanks{Bogdan Marinescu is with École Centrale Nantes, 1 Rue de la Noë, 44000 Nantes, France 
(email: bogdan.marinescu@ec-nantes.fr).}}

\maketitle
\thispagestyle{fancy}

\begin{abstract}

Controller synthesis in power electronics-based systems depends predominantly on the mathematical model of the system, which is a limitation when the actual system is complex and the mathematical model cannot capture all its dynamics. Model-free control addresses this limitation by using an ad-hoc simple model which is compensated by high-rate evaluation of dynamics in terms of their derivatives. However, application of the model-free control strategy to power electronics-based multi-time scale dynamical systems is challenging because of the derivative action needed to implement such control. Grid-connected power converters are examples of such systems, yet experimental validation has not been adequately addressed in the literature. This letter presents the validation of such control including the hardware implementation level. An intelligent proportional–integral (iPI) controller is synthesized and validated on a 16 kW experimental test bench. This proves the  benefits of the approach in control of grid-connected power converters, among which their participation in the secondary voltage control.
 
\end{abstract}


\section{Introduction} \label{section:1}

\IEEEPARstart{C}{onventional} approaches to design controllers for any system are model-based, i.e., the controllers are designed based on a mathematical model of the system. Therefore, their performance is highly dependent on the accuracy of such mathematical model. Many of the real-world systems are complex with some unknown and time-varying dynamics. For example, the controller designed for a power converter connected to a grid may not give the same performance when the grid has a different dynamic behavior. Moreover, it is difficult to capture the switching and nonlinear dynamics of the converter into a simple control model. As a result, construction of a good control model is a challenging task and results in a high-order model. Such a control model is difficult to use for the synthesis of the controller and leads to high-order controllers which cannot be implemented in practice\cite{ref1}. The model-free approach overcomes these difficulties as it uses an ad-hoc simple model which is compensated by high-rate evaluation of dynamics in terms of their derivatives \cite{ref2}. However, application of the model-free control  strategy (called an intelligent proportional–integral (iPI) controller) to multi-time scale dynamical systems is challenging because of their derivative action. In fact, the application of model-free control for complex systems involving multi-time scale dynamics is not reported in the literature. Some examples of these systems are: energy storage \cite{ref3}, lateral vehicle control \cite{ref4}, humanoid robots \cite{ref5}, HVDC systems \cite{ref6}, and voltage control in power systems \cite{ref7}.

\begin{figure}[ht]
    \centering
    \includegraphics[width=1\linewidth]{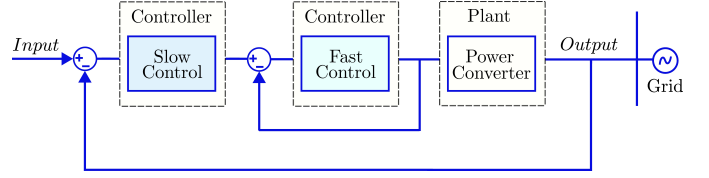}
    \caption{General schematic of multi-time scale control of grid-connected power converter}
    \label{fig:MTSD}
\end{figure}

In this work, we target to explore the feasibility of model-free control for the different systems mentioned above. As an example, we have analyzed a power electronics (PE) based systems involving grid-connected power converters, encountered in the following power systems applications: energy storage, HVDC systems and voltage control. In all these applications, the grid-connected converters have multi-time scale dynamics involving fast inner loops (2–5 ms) and slow outer loops (20–50 ms). The general schematic of these systems is shown in Fig. \ref{fig:MTSD}. This framework can be extended to the application of secondary voltage control (SVC), which was first implemented by Électricité de France (EDF) and has an even broader multi-time scale structure because of a supplementary very-slow grid-voltage loop control with time constant between 200-300s. The reference input of the slow outer loop in Fig. 1 (automatic voltage regulator for the SVC) is further adjusted by another slower controller  to maintain the bus voltages at some selected buses called pilot points \cite{ref7}. While SVC in power systems is traditionally implemented using centralized schemes, a decentralized approach based on a model-free control strategy has been proposed in \cite{ref8}. This decentralized approach is in fact very interesting for renewable sources because centralized SVC requires measurements from few pilot points, which is manageable in conventional power systems but becomes problematic in grids with distributed renewable energy sources. 

Existing work on model-free control has been limited to simulation or scaled-down experimental validation. To the best of the author’s knowledge, model-free control has not yet been implemented or experimentally validated for power systems applications with multi-time-scale dynamics. This letter validates the hardware implementation of model-free control, which is a new control method for grid-connected converters. Intelligent proportional–integral (iPI) controller is synthesized based on the model-free control method for the outer slow control loop that enables the participation of renewable sources in the framework of multi-time scale system. The outer slow control loop is deliberately tuned to a slower time scale to mimic SVC dynamics. This new control method is validated on a 16 kW experimental test bench of a grid-connected power converter.

\section{Controller Synthesis}

\begin{figure}[ht]
    \centering
    \includegraphics[width=0.7\linewidth]{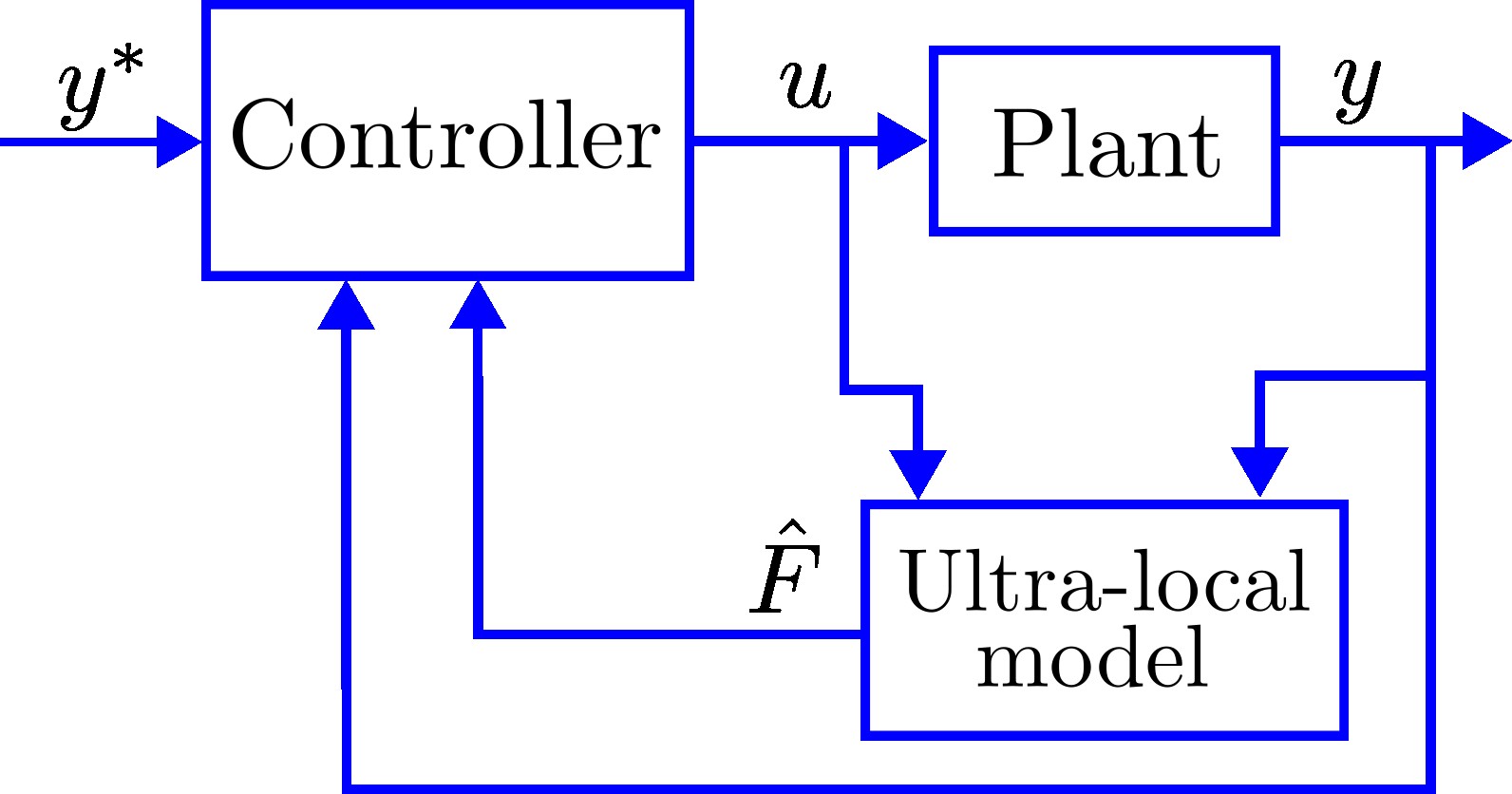}
    \caption{Model-free control scheme}
    \label{fig:MFC}
\end{figure}

The basic idea of model-free control (MFC) is to use as control model an ultra-local first order model, as shown in \eqref{eq:ULM} \cite{ref2}. For simplicity, a single-input single-output system is considered, as illustrated in fig. \ref{fig:MFC}.

\begin{equation}
\dot{y}=F+\alpha u
\label{eq:ULM}
\end{equation}

$\alpha \in \mathbb{R}$ is a non-physical constant chosen so that $\alpha u$ and $\dot{y}$ have the same order of magnitude, $F$ gathers the unknown dynamics and disturbances, and $u$ and $y$ are the input and output of the plant, respectively.

By closing the loop with a controller gives \eqref{eq:IPE1}:

\begin{equation}
u = \underbrace{\frac{\dot{y}^* + K_p e + K_i \int e}{\alpha}}_{\substack{\text{Closed-loop} \\ \text{tracking}}}
- \underbrace{\frac{F}{\alpha}}_{\substack{\text{Unknown dynamics} \\ \text{compensation}}}
\label{eq:IPE1}
\end{equation}

where $e = y - y^*$ is the tracking error, $y^*$ is the reference signal, and $K_p$ and $K_i$ are the controller gains. This is called an intelligent PI (iPI) controller \cite{ref2}. Combining \eqref{eq:ULM} and \eqref{eq:IPE1} gives:

\begin{equation}
\dot{e}+K_pe+K_i\int e=0
\label{eq:CE}
\end{equation}

which ensures local exponential stability for $K_p>0$ and $K_i>0$. For implementation, the online estimation of $F$  is done as:

\begin{equation}
\hat{F}=\dot{y}-\alpha u(t-h)
\label{eq:FEM}
\end{equation}

where $\hat{F}$ is the online estimate of $F$ and $u(t-h)$ is the delayed control input over a short time window. The derivative $\dot{y}$ is calculated based on numerical differentiation techniques to avoid the effect of noisy signals on the controller (see, e.g.,  \cite{ref9}). Therefore, the tuning of $K_p$ and $K_i$ becomes straightforward, which is a major advantage over classical PID controllers.

\begin{figure*}[t!]
\centering
\includegraphics[width=0.75\linewidth]{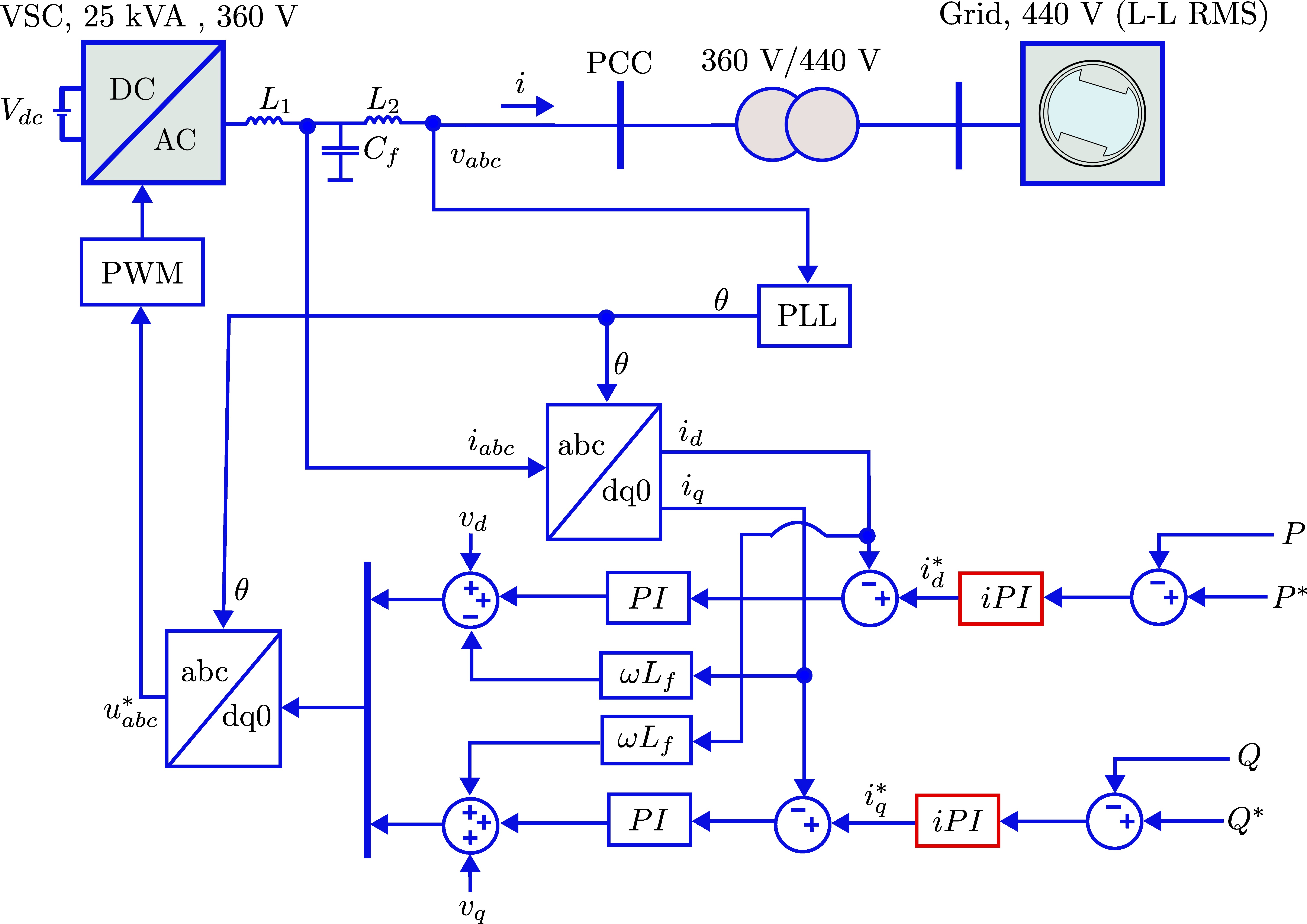}
\caption{System architecture with the iPI controller.}
\label{fig:Sys}
\end{figure*}

Based on this MFC framework, an iPI controller is synthesized for the grid-following converter. The synthesized controller applies only to the slow loop, and the inner loop remains classical PI controller (see, e.g., \cite{ref10}). Fig.~\ref{fig:Sys} shows the control architecture of the grid-following converter with iPI controller. For the outer loop, the dynamics of the inner loop is ignored as it is much faster. Active and reactive powers are directly controlled by the setpoints, and the control objective is to ensure asymptotic tracking of the power references:

\begin{equation}
\lim_{k\to\infty} e_1(k) = 0, \quad \lim_{k\to\infty} e_2(k) = 0
\label{eq:objective}
\end{equation}

where the tracking errors are defined as:

\begin{equation}
e_1(t) = P^*(t) - P(t), \quad e_2(t) = Q^*(t) - Q(t)
\label{eq:errors}
\end{equation}

$P^*(t)$ and $Q^*(t)$ are the desired active and reactive power setpoints, $P(t)$ and $Q(t)$ are the instantaneous powers. Note that the control problem is multivariable as the system has 2 inputs and 2 outputs. Usually in MFC, such situations are simplified by considering a diagonal structure of the regulator \cite{ref2}. More specifically, an independent control law is proposed for each tracking objective. For the control of the active power, the ultra-local model is defined as \eqref{eq:ulm_P}: 

\begin{equation}
\dot{P}(t) = F_1(t) + \alpha_1 u_1(t), \qquad \alpha_1 > 0
\label{eq:ulm_P}
\end{equation}

Here, $u_1(t)$ is the controlled input and $F_1(t)$ is the unknown disturbances. Now by closing the loop with an iPI controller, it gives the following e.q. \ref{eq:control_P}: 

\begin{equation}
u_1(t) = \frac{1}{\alpha_1}\left( \dot{P}^*(t) + K_{p1} e_1(t) + K_{i1} \int_0^t e_1(\tau)\,d\tau - F_1(t) \right)
\label{eq:control_P}
\end{equation}

To ensure asymptotic tracking of the reference, the values of the gains should be positive, i.e., $K_{p1} > 0$ and $K_{i1} > 0$. Substituting the ultra-local model \eqref{eq:ulm_P} into \eqref{eq:control_P} yields the closed-loop error dynamics as:

\begin{equation}
\dot{e}_1(t) + K_{p1} e_1(t) + K_{i1} \int_0^t e_1(\tau)\,d\tau = 0
\end{equation}

In practice, the unknown term $F_1(t)$ is estimated online. From \eqref{eq:ulm_P}, the estimate is obtained as:

\begin{equation}
\hat{F}_1(t) = \dot{P}(t) - \alpha_1 u_1(t-h)
\end{equation}

where $h$ is a small time delay (e.g., one sampling period) and the derivative $\dot{P}$ is computed using numerical differentiation. Substituting $\hat{F}_1(t)$ into the control law in \eqref{eq:control_P} gives the final expression of iPI controller in eq. \ref{eq:iPIP}:

\begin{equation}
u_1(t) = 
\underbrace{\frac{ \dot{P}^*(t) + K_{p1} e_1(t) + K_{i1} \int_0^t e_1(\tau)\,d\tau }{\alpha_1}}_{\substack{\text{Closed-loop} \\ \text{tracking}}}
\;-\;
\underbrace{\frac{ \hat{F}_1(t) }{\alpha_1}}_{\substack{\text{Unknown dynamics} \\ \text{compensation}}}
\label{eq:iPIP}
\end{equation}

The controller synthesis for reactive power is also followed by the similar procedure with $\alpha_2 > 0$ and the gains $K_{p2}, K_{i2} > 0$. This gives the iPI controller expression for reactive power in eq. \ref{eq:iPIQ}:

\begin{equation}
u_2(t) = 
\underbrace{\frac{ \dot{Q}^*(t) + K_{p2} e_2(t) + K_{i2} \int_0^t e_2(\tau)\,d\tau }{\alpha_2}}_{\substack{\text{Closed-loop} \\ \text{tracking}}}
\;-\;
\underbrace{\frac{ \hat{F}_2(t) }{\alpha_2}}_{\substack{\text{Unknown dynamics} \\ \text{compensation}}}
\label{eq:iPIQ}
\end{equation}

with $\hat{F}_2(t) = \dot{Q}(t) - \alpha_2 u_2(t-h)$. The values of the tuning parameters $\alpha_1, \alpha_2, K_{p1}, K_{p2}, K_{i1}, K_{i2}$  are given in table \ref{tab:PBMS}. The gains are tuned to achieve the dynamics of SVC, i.e., 200 s, as explained in section \ref{section:1}.

\section{Experimental Validation}

\begin{figure}
    \centering
    \includegraphics[width=1\linewidth]{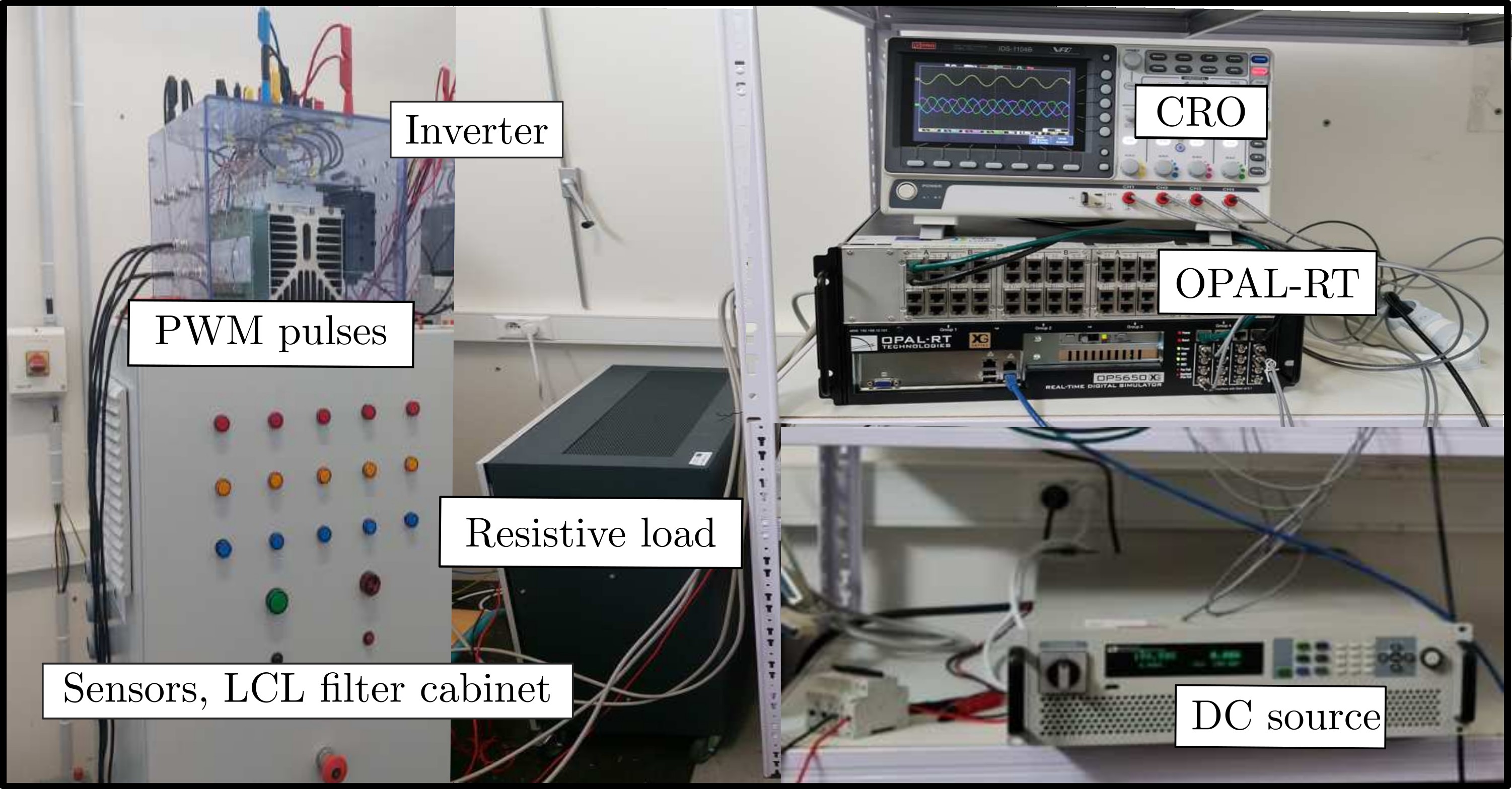}
    \caption{Benchmark experimental setup}
    \label{fig:BSR}
\end{figure}

The experimental benchmark system consists of a three-phase converter with a nominal active power rating of 16 kW. Fig.~\ref{fig:BSR} shows the experimental setup. The converter is synchronized to the grid via a phase-locked loop (PLL). Line-to-line RMS voltage is 440 V, and the fundamental frequency is 50 Hz. A 360/440 V step-down transformer is used to connect the grid to the converter. The converter is connected to the grid through an LCL filter, and DC-link voltage is regulated by a constant DC source. In addition, a resistive load is connected at the point of common coupling. Detailed parameters are listed in Table~\ref{tab:PBMS}. Real-time control and pulse-width modulation (PWM) signal generation are implemented using an OPAL-RT OP5650 real-time simulator. The embedded Artix-7 FPGA generates PWM signals at a switching frequency of 5 kHz, which are applied to the converter through high-speed digital I/O channels. The sampling time period of both the system and controller is $T_s = 50~\mu$s.  The experimental setup represents a medium power grid-connected converter commonly used in distributed energy resource applications, such as photovoltaic and small-scale wind energy systems.

\begin{table}[htbp]
\centering
\caption{Parameters of the benchmark system}
\label{tab:PBMS}
\begin{tabular}{|l|l|}
\hline
\textbf{Parameter} & \textbf{Value} \\
\hline
Grid voltage (RMS), $V_{grid}$ & 440 V \\
\hline
DC link voltage, $V_{DC}$ & 750 V \\
\hline
Active power, $P$ & 16 kW \\
\hline
Filter inductance, $L_1$, $L_2$ & 1.18 mH, 0.618 mH \\
\hline
Filter resistances, $R_1$, $R_2$ & 0.1 $\Omega$ \\
\hline
Filter capacitance, $C_f$ & 60 $\mu$F \\
\hline
Grid frequency, $f$ & 50 Hz \\
\hline
Switching frequency, $f_s$ & 5 kHz \\
\hline
Controller sampling time, $T_s$ & 50 $\mu$s \\
\hline
PWM sampling time, $T_s$ & 1 $\mu$s \\
\hline
Resistive load, $R_l$ & 4 kW \\
\hline
Non-physical constant, $\alpha_1, \alpha_2$ & $1/8000$ \\
\hline
Proportional gain, $K_{p1}=K_{p2}$ & $0.002$ \\
\hline
Integral gain, $K_{i1}=K_{i2}$ & $0.0009$ \\
\hline
\end{tabular}
\end{table}

\begin{figure}
    \centering
    \includegraphics[width=1\linewidth]{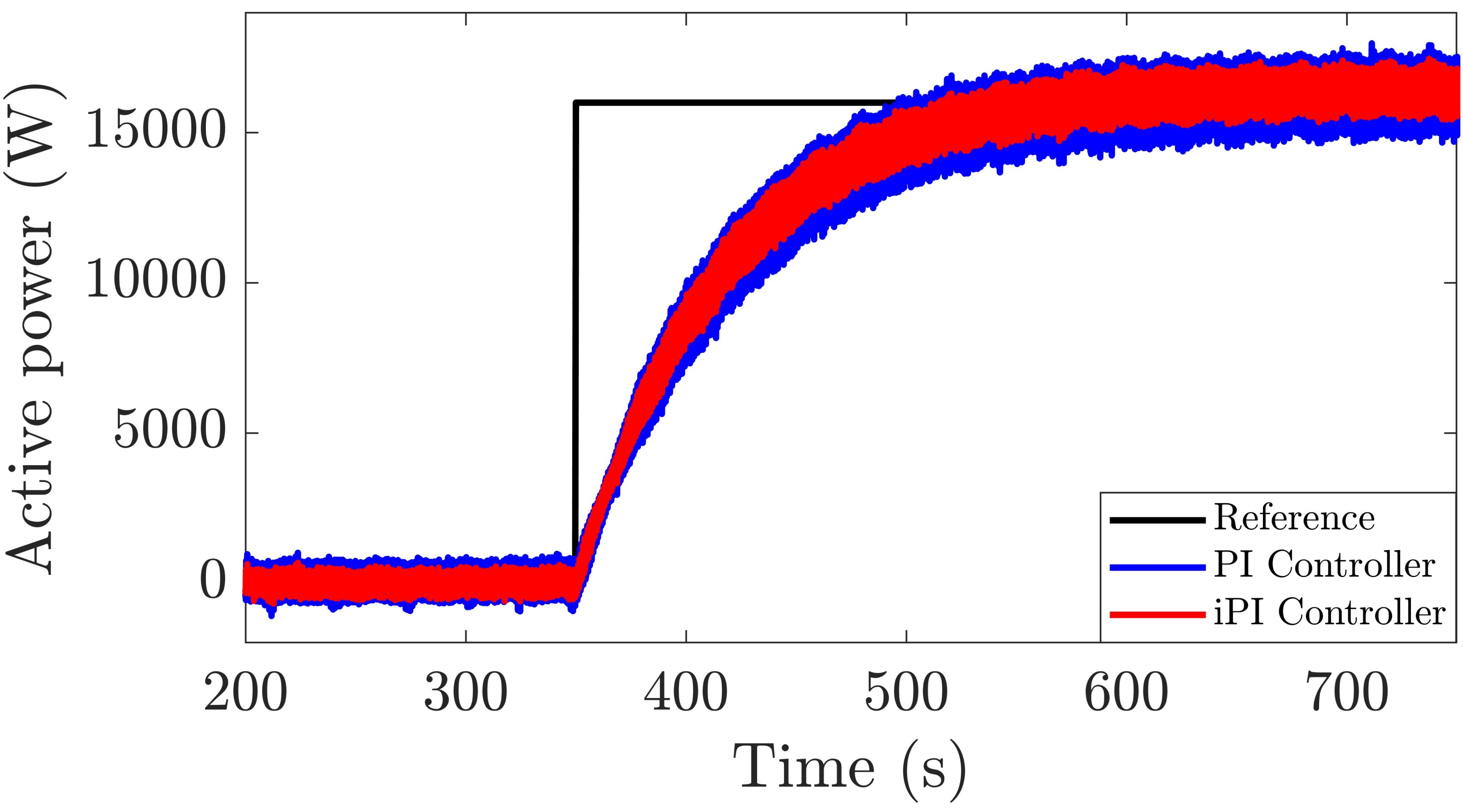}
    \caption{Performance comparison for a step change of active power.}
    \label{fig:PC}
\end{figure}

\begin{figure}
    \centering
    \includegraphics[width=1\linewidth]{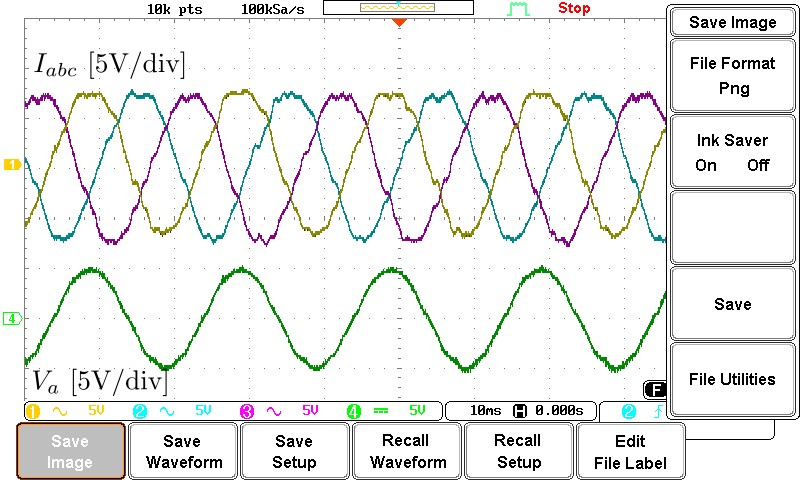}
    \caption{Voltage and current waveforms in oscilloscope for 16~kW active power.}
    \label{fig:SP}
\end{figure}

Fig. \ref{fig:PC} illustrates the dynamic performance of the grid-connected converter for a step change of 16~kW in active power. The performance of the iPI controller is validated against the standard PI controller. To have a fair comparison, both of the controllers are tuned as per the timescale of SVC dynamics. A step change in active power reference is applied at 350 s, and the system response settles around 550 s. This validates that the desired dynamics with a time constant around 200 s mentioned in section II is achievable with the iPI controller. Fig.~\ref{fig:SP} shows the voltage and current waveform of the iPI controller at 16~kW active power. Three-phase current and single-phase voltage are shown in the oscilloscope with the scaling of 5 V/div. Importantly, the reconstructed derivative does not introduce any additional noise or degrade the waveforms. Also, the obtained total harmonic distortion (THD) is 4.31\%. The iPI controller satisfies the harmonic standard (IEEE 1547) as THD below 5\%, whereas the PI controller has a THD of 6.2\%, which is above the standards and thus not implementable. Hence, it can be concluded that the iPI controller is suitable for power systems applications even though a derivative term is present. 

\section{Conclusion}
This letter proved the feasibility of a model-free control on grid connected converters as systems with multi-time scale dynamics.  For that, an iPI controller was synthesized and validated on an experimental benchmark of 16 kW which mimics the dynamics of the secondary voltage control. It was observed that the numerical differentiation involved in the controller did not introduce any additional noise or degrade the waveforms, thus making it suitable for all grid-connected power system applications which present the same kind of multi-time scale dynamics. Moreover,  this work could serve as a foundation for the real-world applications of model-free control in other domains involving multi-time-scale dynamics (like, e.g., vehicle control, humanoids robots, trajectory tracking).


\begin{thebibliography}{1}
\bibliographystyle{IEEEtran}

\bibitem{ref1}
W. Li, H. Yuan, S. Li, and J. Zhu, ``A revisit to model-free control,'' \textit{IEEE Trans. Power Electron.}, vol. 37, no. 12, pp. 14408--14421, 2022.

\bibitem{ref2}
M. Fliess and C. Join, ``Model-free control,'' \textit{Int. J. Control}, vol. 86, no. 12, pp. 2228--2252, 2013.

\bibitem{ref3}
Y. Hong, D. Xu, W. Yang, B. Jiang, and X.-G. Yan, ``A novel multi-agent model-free control for state-of-charge balancing between distributed battery energy storage systems,'' \textit{IEEE Trans. Emerg. Topics Comput. Intell.}, vol. 5, no. 4, pp. 679--688, 2020.

\bibitem{ref4}
L. Menhour, B. d'Andr{\'e}a Novel, M. Fliess, D. Gruyer, and H. Mounier, ``An efficient model-free setting for longitudinal and lateral vehicle control: Validation through the interconnected pro-sivic/rtmaps prototyping platform,'' \textit{IEEE Trans. Intell. Transp. Syst.}, vol. 19, no. 2, pp. 461--475, 2017.

\bibitem{ref5}
J. Villagra and C. Balaguer, ``Robust motion control for humanoid robot flexible joints,'' in \textit{Proc. 18th Mediterr. Conf. Control Autom. (MED)}, Marrakech, Morocco, 2010, pp. 963--968.

\bibitem{ref6}
A. Barkat, B. Marinescu, C. Join, and M. Fliess, ``Model-free control for VSC-based HVDC systems,'' in \textit{Proc. IEEE PES Innov. Smart Grid Technol. Conf. Eur. (ISGT-Europe)}, Sarajevo, Bosnia and Herzegovina, 2018, pp. 1--6.

\bibitem{ref7}
B. Marinescu, ``Robustness and coordination in voltage control of large-scale power systems,'' \textit{Int. J. Control}, vol. 81, no. 10, pp. 1568--1589, Oct. 2008.

\bibitem{ref8}
J. K. Goyal, V. Thomas, and B. Marinescu, ``A decentralised control strategy for secondary voltage regulation,'' arXiv:2205.13833, 2022.

\bibitem{ref9}
M. Mboup, C. Join, and M. Fliess, ``Numerical differentiation with annihilators in noisy environment,'' \textit{Numer. Algorithms}, vol. 50, no. 4, pp. 439--467, 2009.

\bibitem{ref10}
W. Leonhard, \textit{Control of Electrical Drives}. Berlin, Germany: Springer, 2001.

\end{thebibliography}
\end{document}